\documentclass[11pt]{article}

\usepackage[preprint]{acl}

\usepackage{times}
\usepackage{latexsym}

\usepackage{booktabs}
\usepackage{multirow}
\usepackage{graphicx}
\usepackage[table]{xcolor}
\usepackage{amsmath,tabularx,array}
\usepackage[most]{tcolorbox}
\usepackage[T1]{fontenc}
\usepackage{adjustbox}
\usepackage{xspace}
\usepackage{enumitem}

\definecolor{PromptBlue}{HTML}{5977AB}
\definecolor{PromptBg}{HTML}{F2F6FA}

\newcommand{\ours}{\textsc{Crossfeed}\xspace}

\usepackage[T1]{fontenc}

\usepackage[utf8]{inputenc}

\usepackage{microtype}

\usepackage{inconsolata}

\usepackage{graphicx}

\title{\ours: Self-Improving Unified Audio Models \\ across Understanding, Judging, and Generation with Zero External Data}

\author{
  Yu-Xiang Lin \\
  {\normalfont National Taiwan University} \\
  \\
  {\normalfont\ttfamily f13942075@ntu.edu.tw}
  \And
  Hung-yi Lee \\
  {\normalfont National Taiwan University} \\
  {\normalfont NTU AI-CoRE} \\
  {\normalfont\ttfamily hungyilee@ntu.edu.tw}
}

\begin{document}
\maketitle


\begin{abstract}
    Recent post-training methods have explored ways to improve the audio understanding of large audio language models, but they typically rely on real audio corpora or external models such as audio synthesizers, teachers, and judges.
    The ability of unified audio models to generate both text and audio raises a natural question:~\textit{can these models generate their own training data and use it to improve themselves?}
    We introduce \textsc{Crossfeed}, a post-training framework that improves the understanding, judging, and generation capabilities of a unified audio model with zero external data and no teacher models. 
    In this framework, the model synthesizes non-speech sound and speech from self-generated captions, transcripts, and dialogue scripts. 
    By constructing labeled data and preference pairs from its own outputs, the model trains itself to improve all three capabilities.
    We also find that self-merging, which retains half of each fine-tuning update, leads to more consistent improvements across model capabilities.
    Experiments on UNISON and Audex-2B show that \textsc{Crossfeed} improves performance on most benchmarks spanning understanding, judging, and generation.
\end{abstract}

\begin{figure}
    \centering
    \includegraphics[width=\linewidth]{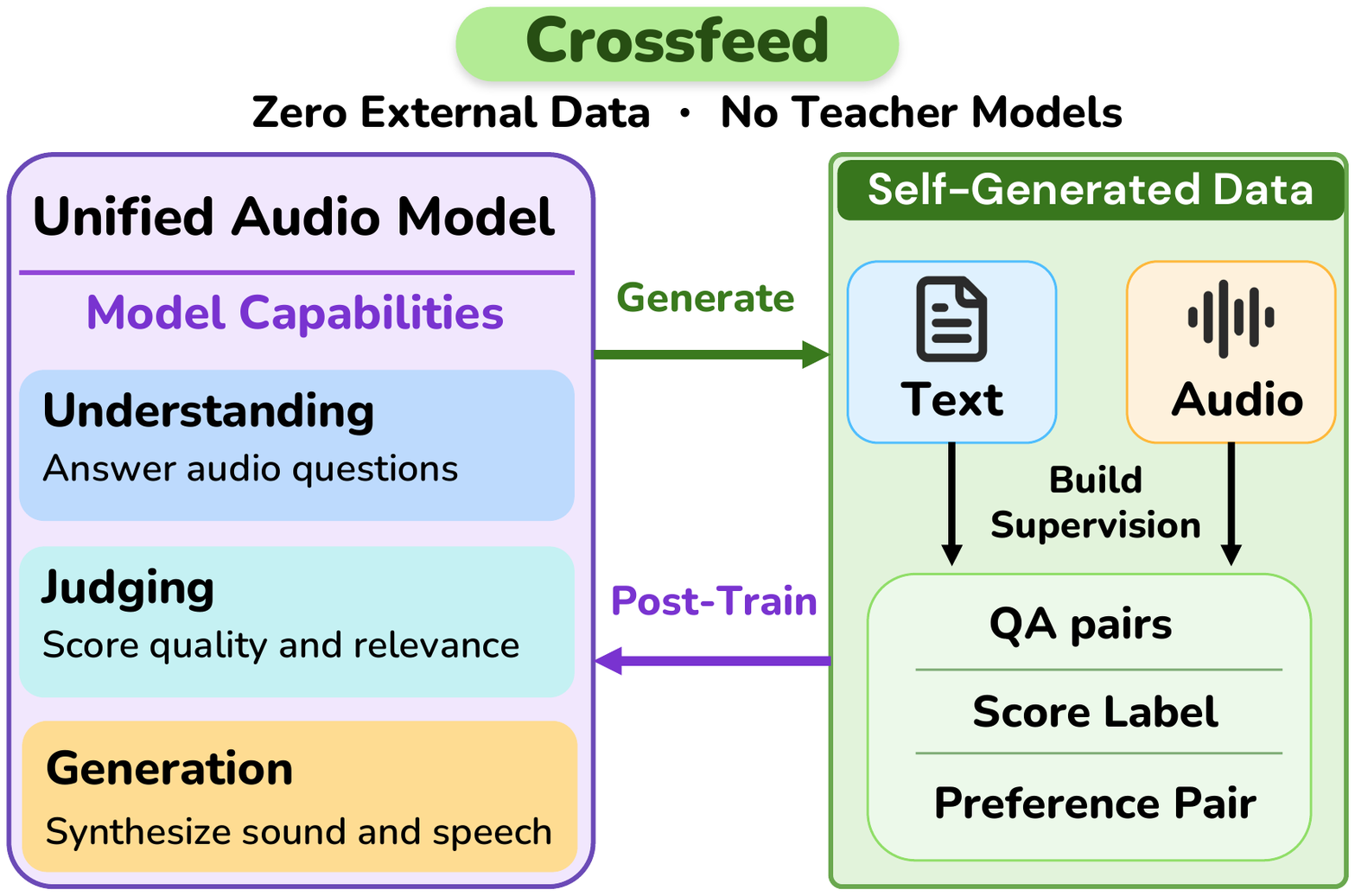}
    \caption{Overview of~\ours. ~\textit{Generation}, ~\textit{understanding}, and~\textit{judging} each contribute to training data construction, and their outputs jointly supervise the model itself without external data or teacher models.}

    \label{fig:placeholder}
\end{figure}

\section{Introduction}
Unified multimodal models (UMMs) have emerged as a promising direction for integrating multimodal understanding and generation within a shared modeling framework. 
In the audio domain, recent UMMs\footnote{While spoken language models (SLMs) can understand both speech and non-speech sound, their generation capabilities primarily focus on speech. 
In this work, we focus on UMMs that additionally support non-speech sound generation.} support speech and general audio understanding and generation~\cite{anygpt,ualm,audex,audio-omni,unison}.
Beyond supporting multiple tasks within a single model, such unification opens up new opportunities for self-improvement.

Recent studies have explored post-training methods to improve the audio understanding of large audio language models (LALMs).
These methods typically train on real audio corpora, with supervision from human annotations~\citep{r1-aqa, audiothinker} or external teacher models~\citep{audio-reasoner}.
Even self-improving methods that require no external supervision still rely on existing audio data~\citep{si-sda, aqa-ttrl, audio-zero} and focus only on improving understanding.
With both understanding and generation capabilities, a unified audio model can generate its own training data.
This raises a natural question:~\textit{Can a unified audio model improve its understanding, judging, and generation without external data or models?}

In this work, we introduce a \textit{self-improving post-training} framework in which an off-the-shelf audio UMM leverages its existing capabilities to improve its understanding, judging, and generation without external training data or teacher models.
First, the model synthesizes audio and speech with its generation capability.
It then labels these samples itself, proposing and answering questions to train understanding and producing quality and relevance scores to train judging.
Next, the model scores its own generated samples, and the resulting preference pairs are used to train generation.
Finally, we apply self-merging by adding half of each fine-tuning update to the original model weights to incorporate newly learned capabilities while better retaining its original capabilities.
Our contributions are summarized as follows:

\begin{itemize}[leftmargin=16pt, itemsep=0pt, topsep=3pt, after=\vspace{3pt}, labelsep=0.5em]
    \item We introduce a self-improving post-training framework,~\textbf{\ours}\footnote{Our code is publicly available at~\url{https://github.com/YuXiangLin1234/Crossfeed}}, that leverages the existing capabilities of audio UMMs to generate their own supervision, without relying on external training data or teacher models.
    \item We evaluate \ours~on two audio UMMs, UNISON and Audex-2B, and observe improvements on most benchmarks spanning understanding, judging, and generation.
    \item We find that self-merging helps reduce performance regressions and yields more consistent gains across benchmarks.
\end{itemize}

\begin{figure*}
    \centering
    \includegraphics[width=\textwidth]{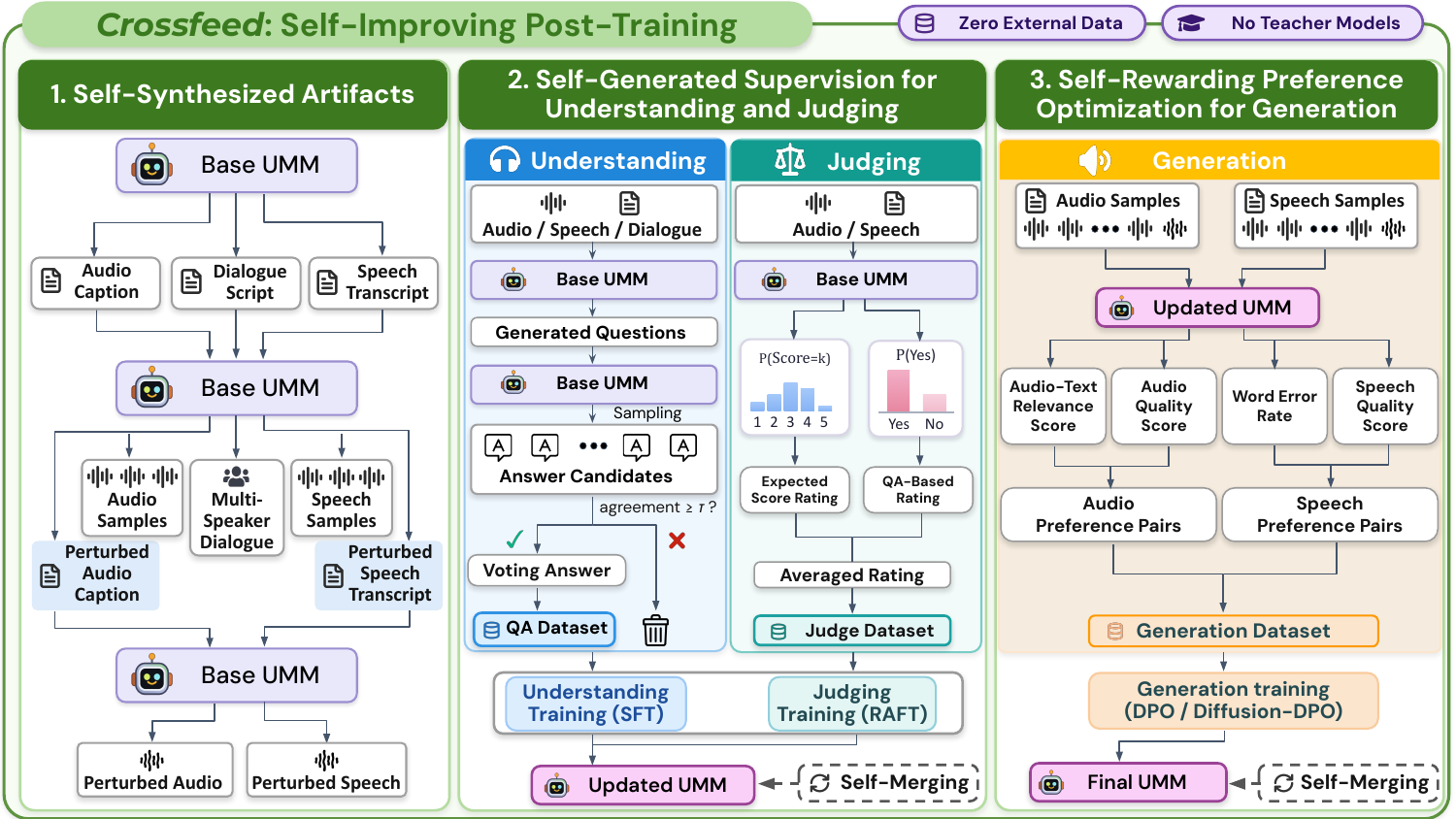}
    \caption{Overview of the~\ours training framework.}
    \label{fig:method_overview}
\end{figure*}

\section{Related Work}

\subsection{Unified Models for Speech and Audio Understanding and Generation}

LALMs enable large language models (LLMs) to understand speech and non-speech audio~\cite{salmonn,mimoaudio,desta2.5}. 
A common design feeds representations from a pretrained audio encoder, such as Whisper~\cite{whisper}, into an LLM backbone. 
LALMs typically respond in text, and some produce spoken responses through a TTS module~\cite{audioflamingo3} or a Talker~\cite{qwen25omni}.

Unified audio models combine these understanding capabilities with the ability to generate both speech and non-speech audio.
Some models use a shared autoregressive LLM for understanding and generation.
AnyGPT represents text, speech, images, and music as discrete tokens at the input~\citep{anygpt}.
Audex feeds continuous audio embeddings into the LLM and generates discrete tokens for speech and non-speech audio~\citep{audex}.
Other models use an LALM to condition a diffusion transformer (DiT)~\cite{dit} for audio generation.
Audio-Omni conditions its DiT on hidden states from the penultimate LALM layer~\citep{audio-omni}, while UNISON uses hidden states from multiple LALM layers to condition corresponding DiT blocks~\citep{unison}.
In this work, we study Audex and UNISON to cover both architectures.

\subsection{Self-Improving Models}
Self-improvement aims to enable models to enhance their own capabilities using self-generated training signals. 
This paradigm has been widely studied in LLMs.
~\citet{self-rewarding} use the model itself as a judge to reward its own responses during iterative DPO training.
Absolute Zero~\cite{absolute-zero} and R-Zero~\cite{r-zero} further remove the need for external data by letting a base model propose its own tasks and learn to solve them.
Similar ideas have been extended to vision~\cite{unirl,srum,unicorn}.
Most closely related to our work, UniCorn~\cite{unicorn} uses self-generated data to improve generation while preserving its understanding capabilities.
In contrast, our work aims to improve understanding, judging, and generation within a single framework.

In the speech domain, previous work used separate recognition and synthesis models to provide training signals for each other~\cite{speech-chain,token-chain}.
Other studies have explored how LALMs can improve themselves on audio-to-text tasks.
Audio-Zero~\cite{audio-zero} introduces a self-play framework based on the game~\textit{Who Is the Spy}, removing the need for explicit audio QA data but still requiring audio pairs with similar content and subtle acoustic differences.
Other approaches further relax the requirements on the training data. 
SI-SDA~\cite{si-sda} uses information stored in model attention to learn from unlabeled speech.
AQA-TTRL~\cite{aqa-ttrl} performs test-time adaptation directly on unlabeled test data to improve LALM performance.
These methods remove the need for paired audio samples, but still rely on real audio corpora and focus only on improving understanding.

Our work explores self-improvement in audio UMMs without external training data.
Given a pretrained UMM, we use task-specific prompts to generate training data and derive supervision from the model itself to improve its understanding, judging, and generation capabilities.
External datasets and models are used only for evaluation.

\section{Method}
\label{sec:method}
Our goal is to improve an audio UMM by explicitly turning its own outputs into training signals.
Figure~\ref{fig:method_overview} illustrates this process.
The audio UMM first acts as a~\textit{generator}, producing text instructions and synthesizing the corresponding audio and speech (\S\ref{subsec:method_artifacts}).
The generated audio then provides the basis for understanding and judging data.
As a~\textit{QA proposer}, it proposes and answers questions according to the audio.
As a~\textit{judge}, it evaluates the audio, and we derive pseudo-labels from its output probabilities.
We fine-tune the base UMM on both datasets (\S\ref{subsec:method_understanding_judging}).
The updated UMM then takes on the judging role again and scores the generated samples to construct preference pairs for generation training via DPO~\cite{dpo} (\S\ref{subsec:method_preference_optimization}).
We also apply model merging at selected points during post-training to integrate the learned updates and obtain the final model (\S\ref{subsec:method_merge}).
Appendix~\ref{app:implementation_details} provides additional implementation details.

\subsection{Self-Synthesized Artifacts}
\label{subsec:method_artifacts}
 
\paragraph{Instruction and Candidate Generation.}
We first prompt the model to generate two types of text instructions: audio captions describing general sound scenes and utterance transcripts for speech synthesis.
For each instruction, the model synthesizes $N$ audio candidates through Text-to-Audio (TTA) or TTS generation.
These candidates provide diverse realizations of the same instruction, which are later used to construct preference pairs for generation optimization (\S\ref{subsec:method_preference_optimization}).
 
\paragraph{Perturbation-Based Negative Generation.}
To construct semantic negative samples, we prompt the model to make small changes to the original instruction $c$.
For audio captions, the model is instructed to modify event order, timing, repetition counts, sound attributes, or the presence of sound events.
For speech transcripts, it is instructed to substitute phonetically similar words or alter word repetition counts while preserving the rest of the utterance.
These modifications yield perturbed instructions $\tilde{c}$ that closely resemble $c$ but describe different target content.
The model then synthesizes the audio $\tilde{a}$ conditioned on $\tilde{c}$.
We use $\tilde{a}$ as a negative candidate for $c$ and as a candidate response to $\tilde{c}$ when constructing instruction-conditioned preference pairs (\S\ref{subsec:method_preference_optimization}).
 
\paragraph{Multi-speaker dialogues.}
We also prompt the model to generate multi-speaker scripts for dialogue understanding.
Each script specifies the speaker and utterance for each turn.
The model then synthesizes the corresponding multi-speaker speech.
We use these generated speech samples and the corresponding scripts to construct QA pairs about speaker identities, turn order, and dialogue content (\S\ref{subsec:method_understanding_judging}).

\subsection{Self-Generated Supervision for Understanding and Judging}
\label{subsec:method_understanding_judging}

We construct question-answering (QA) data and pseudo-labeled scores from self-generated audio artifacts to improve the UMM's understanding and judging capabilities, respectively.

\paragraph{Question Generation.}
Given an audio sample and its corresponding caption, transcript, or dialogue script, the model generates multiple questions about the audio.
For multi-speaker dialogues, these questions also cover speaker identities, turn order, and dialogue content.

\paragraph{Agreement-Based QA Verification.}
We adopt an agreement-based verification strategy based on self-consistency~\cite{self-consistency} to filter unreliable QA pairs.
For each question, the answerer samples $K$ responses, and we retain the question only if the most frequent answer accounts for at least a fraction $\tau$ of the responses.
The most frequent answer is then used as the final answer.

\paragraph{Probability-Based Judging Labels.}
For judging, we construct pseudo labels for three assessment tasks: audio-text relevance and audio quality for TTA outputs, and speech quality for TTS outputs.

We use two probability-based scoring approaches.
~\textit{Score-expectation-based scoring} computes a probability-weighted average over integer ratings from 1 to 5, based on G-Eval~\cite{g-eval}, which was originally proposed for text generation evaluation.
~\textit{QA-based scoring} formulates evaluation as a binary yes/no question and uses the normalized probability of answering ``Yes'' as the score, based on AQAScore~\cite{aqascore}.
While AQAScore was originally developed for audio-text relevance, we extend this formulation to audio and speech quality assessment by adapting the evaluation questions.

We denote the scores from~\textit{score-expectation-based scoring} and~\textit{QA-based scoring} as
$\hat{y}_{\mathrm{SE}}$ and $\hat{y}_{\mathrm{QA}}$, respectively.
We first linearly transform the QA-based score from $[0,1]$ to the $[1,5]$ scale:
\begin{equation}
    \tilde{y}_{\mathrm{QA}} = 4\hat{y}_{\mathrm{QA}} + 1.
\end{equation}

We then average the two scores to obtain the final pseudo label:
\begin{equation}
\label{eq:judge-score}
\hat{y} = \frac{\hat{y}_{\mathrm{SE}} + \tilde{y}_{\mathrm{QA}}}{2}.
\end{equation}

\paragraph{Joint Understanding and Judging Training.}
With both training datasets constructed, we fine-tune the model using separate objectives for understanding and judging.
For understanding, we apply standard supervised fine-tuning (SFT) with cross-entropy loss.
For judging, we adopt regression-aware fine-tuning (RAFT)~\cite{raft}, which minimizes the squared error between the model's probability-weighted predicted scores and the pseudo labels.
We jointly optimize both objectives during training.

\subsection{Self-Rewarding Preference Optimization for Audio and Speech Generation}
\label{subsec:method_preference_optimization}
 
To improve audio and speech generation, we use the model to score the generated audio samples and construct preference pairs for subsequent preference optimization.

\paragraph{Preference Pairs Construction.}
For both TTS and TTA generation, we select preference pairs from candidates generated from the same instruction based on~\textit{Pareto dominance}.
We prefer one candidate over another if it is at least as good on both criteria and the improvement on at least one criterion meets a minimum threshold $\delta$.
For TTS, we consider transcript consistency and speech quality.
We use the model to transcribe each speech sample and measure its word error rate (WER) against the target transcript.
For TTA, we consider audio-text relevance and audio quality.


 
\paragraph{Instruction-Perturbed Negative Pairs.}
We further construct preference pairs using the negative samples generated in \S\ref{subsec:method_artifacts}.
For each original instruction $c$ and its perturbed version $\tilde{c}$, we pair a selected preferred response to $c$ with the audio generated from $\tilde{c}$ as the rejected response, without imposing additional score-margin requirements.
We also construct a mirrored pair by reversing the preferred and rejected responses under $\tilde{c}$.
These pairs encourage the generator to condition its preferences on the given instruction rather than favoring one audio sample regardless of the requested content.

\paragraph{Preference Optimization.}
We collect preference pairs into the generation training dataset and optimize the model using DPO~\cite{dpo}.
For autoregressive UMMs, we apply DPO to the likelihoods of discrete audio tokens.
For diffusion-based UMMs, we adopt Diffusion-DPO~\cite{diffusion-dpo}, which uses denoising errors to approximate the preference optimization objective.

\subsection{Self-Merging}
\label{subsec:method_merge}

We use model merging to incorporate what the model learns during fine-tuning while preserving the robustness of the pretrained model.
Prior work has shown that weight interpolation can improve robustness under distribution shifts in contrastive vision-language models~\cite{wise-ft}, preserve safety alignment and instruction-following capabilities in fine-tuned LLMs~\cite{safeguard-merging}, and combine task-specific updates through task arithmetic~\cite{task-arithmetic}. 

Inspired by these findings, we use \emph{self-merging} to keep half of each fine-tuning update relative to the original pretrained weights. 
We fix the merging weight at 0.5 to avoid tuning it on external validation data.
We apply the same rule to both types of models, but the merging procedure depends on the model architecture, as described below.

\paragraph{Diffusion-Based Models.}
For models combining an LALM backbone with a diffusion-based audio generator, such as UNISON~\cite{unison}, we first merge the LALM fine-tuned for understanding and judging with its original weights.
We then fine-tune the DiT for generation using the merged LALM as conditioning.
As in the original training setup, the LALM remains frozen during DiT fine-tuning.
Finally, we merge the fine-tuned DiT with its original weights.

\paragraph{Unified Autoregressive Models.}
For unified autoregressive models like Audex~\cite{audex}, we independently fine-tune the base model for understanding and judging, and for generation.
We then average the two resulting checkpoints with equal weights. 
Each checkpoint consists of the same pretrained weights plus its own fine-tuning update. 
Averaging takes half of both terms from each checkpoint: the two halves of the shared pretrained weights sum to the original pretrained weights, and each fine-tuning update contributes exactly half of its original value.

\section{Experiments}
\subsection{Experimental Setup}
\label{subsec: experimental_setup}
We conduct experiments on two unified audio models: UNISON~\cite{unison} and Audex-2B~\cite{audex}.
UNISON conditions the MM-DiT blocks on the Qwen2.5-Omni-7B thinker~\cite{qwen25omni}, while Audex uses a shared LLM backbone with modality-dedicated encoder and decoder. 
These distinct architectures allow us to evaluate the effectiveness of our method across different unified audio modeling frameworks.
We train our model with LoRA~\cite{lora}. 

Unless otherwise specified, we generate a total of 5,000 samples comprising audio captions, speech transcripts, and dialogue scripts.
We then synthesize the corresponding audio and speech samples and construct task-specific training data from the generated samples.
All experiments are conducted on a single NVIDIA H200 GPU.
Further implementation details are provided in Appendix~\ref{app:implementation_details}.

\subsection{Evaluation Datasets and Metrics}

We evaluate three capabilities:~\textit{understanding}, ~\textit{judging}, and~\textit{generation}.
For understanding, we report accuracy on MMAU~\cite{mmau}, MMAR~\cite{mmar}, MMSU~\cite{mmsu}, and TREA~\cite{trea}.
For judging, we report Pearson correlation (PCC), Spearman correlation (SRCC), and mean squared error (MSE) to assess audio-text relevance on PAM~\cite{pam}, Clotho~\cite{clotho-mos}, XACLE~\cite{xacle}, and RELATE~\cite{relate}; 
audio quality on PAM and Clotho;
and speech quality on BVCC~\cite{bvcc} and QualiSpeech~\cite{qualispeech}. 
For generation, we evaluate TTA on AudioCaps~\cite{audiocaps} using distributional and semantic metrics, and on AudioTime~\cite{audiotime} using temporal metrics. 
We further assess both benchmarks using CoTTAGE~\cite{cottage}, a judge model to evaluate audio-text relevance and audio quality.
For TTS, we report WER and UTMOS on Seed-TTS-Eval~\cite{seed-tts}, CV3-Eval~\cite{cosyvoice3}, and LibriTTS~\cite{libritts}.
Following CV3-Eval, we also report speaker similarity  using ERes2Net~\cite{eres2net} on benchmarks with predefined reference-target pairs.
Further details are provided in Appendix~\ref{app:benchmarks}.

\subsection{Evaluated Methods}
\label{subsec:baselines}

We compare our method against the base models and several baselines.
For understanding, we additionally evaluate self-consistency (SC)~\cite{self-consistency}, which samples eight chain-of-thought reasoning paths~\cite{cot} and selects the majority answer.
For judging, we consider averaged sampling (avg@8), G-Eval-based scoring~\cite{g-eval}, and AQAScore-based scoring~\cite{aqascore}.
To investigate the effect of self-merging, we also evaluate variants of both models with all merging steps omitted.


\providecommand{\betterbg}[1]{\cellcolor{green!20}#1}
\begin{table}[t]
\centering
\begingroup
\scriptsize
\setlength{\tabcolsep}{3pt}
\renewcommand{\arraystretch}{1.05}
\begin{adjustbox}{width=\columnwidth}
\begin{tabular}{lc@{\hspace{8pt}}c@{\hspace{8pt}}c@{\hspace{8pt}}c@{\hspace{8pt}}c@{\hspace{8pt}}c}
\toprule
\textbf{Method}
& \multicolumn{2}{c}{\textbf{MMAU}}
& \textbf{MMAR}
& \textbf{MMSU}
& \textbf{TREA}
& \textbf{Overall} \\
\cmidrule(lr){2-3}
& \tiny Test-mini & \tiny Test &  &  &  &  \\
\midrule
\multicolumn{7}{c}{\textbf{UNISON}} \\
\midrule
Base (greedy) & 74.40 & 73.62 & \textbf{63.10} & 63.46 & \underline{66.67} & \underline{66.73} \\
Base (SC) & \underline{76.00} & \underline{75.10} & 59.70 & \underline{65.36} & 62.83 & 65.77 \\
\addlinespace[2pt]
\rowcolor{gray!10}
\ours & \betterbg{\textbf{77.00}} & \betterbg{\textbf{75.19}} & \underline{63.00} & \betterbg{\textbf{66.44}} & \betterbg{\textbf{71.50}} & \betterbg{\textbf{69.08}} \\
\rowcolor{gray!10}
\quad w/o self-merging & 75.50 & 73.41 & 60.60 & 65.34 & 65.00 & 66.14 \\
\midrule
\multicolumn{7}{c}{\textbf{Audex-2B}} \\
\midrule
Base (greedy) & 70.60 & 71.83 & 51.80 & 57.32 & 56.67 & 59.37 \\
Base (SC) & \textbf{74.10} & \underline{73.08} & 51.30 & \textbf{64.14} & 49.33 & 59.49 \\
\addlinespace[2pt]
\rowcolor{gray!10}
\ours & 73.70 & \betterbg{\textbf{73.56}} & \betterbg{\textbf{55.10}} & 62.10 & \betterbg{\textbf{62.83}} & \betterbg{\underline{63.40}} \\
\rowcolor{gray!10}
\quad w/o self-merging & \betterbg{\textbf{74.10}} & 73.00 & \betterbg{\underline{54.70}} & \underline{63.14} & \betterbg{\textbf{62.83}} & \betterbg{\textbf{63.44}} \\
\bottomrule
\end{tabular}
\end{adjustbox}
\endgroup
\caption{Accuracy on audio understanding and reasoning benchmarks (\%).
Overall is the average across four benchmarks, with MMAU Test-mini and Test combined.
\textbf{Bold} indicates the best performance within each model and column,
while \underline{underline} indicates the second-best.
Gray shading highlights self-training, while green shading indicates results that are better than or equal to all prompting-based and test-time compute baselines.}
\label{tab:understanding}
\end{table}

\providecommand{\betterbg}[1]{\cellcolor{green!20}#1}
\begin{table*}[t]
\centering
\begingroup
\fontsize{8}{9.6}\selectfont
\setlength{\tabcolsep}{4pt}
\renewcommand{\arraystretch}{0.95}
\begin{tabularx}{1\textwidth}{l*{3}{>{\centering\arraybackslash}X}@{\hspace{12pt}}*{3}{>{\centering\arraybackslash}X}@{\hspace{12pt}}*{3}{>{\centering\arraybackslash}X}}
\toprule
\textbf{Method}
& \multicolumn{3}{c}{\textbf{Audio-REL}}
& \multicolumn{3}{c}{\textbf{Audio-OVL}}
& \multicolumn{3}{c}{\textbf{Speech-OVL}} \\
\cmidrule(lr){2-4} \cmidrule(lr){5-7} \cmidrule(l){8-10}
& PCC$\uparrow$ & SRCC$\uparrow$ & MSE$\downarrow$
& PCC$\uparrow$ & SRCC$\uparrow$ & MSE$\downarrow$
& PCC$\uparrow$ & SRCC$\uparrow$ & MSE$\downarrow$ \\
\midrule
\multicolumn{10}{c}{\textbf{UNISON}} \\
\midrule
Prompting (greedy)
& 0.398 & 0.336 & 1.186
& 0.281 & 0.258 & 1.212
& 0.195 & 0.201 & 2.767 \\
Prompting (avg@8)
& 0.455 & 0.409 & 0.921
& 0.296 & 0.269 & 0.736
& 0.277 & 0.252 & 2.086 \\
AQAScore
& 0.536 & \textbf{0.560} & 1.105
& 0.355 & 0.361 & 0.713
& 0.386 & 0.409 & 1.151 \\
G-Eval
& 0.486 & 0.446 & 0.722
& 0.349 & 0.328 & 0.768
& 0.343 & 0.323 & \underline{1.036} \\
AQAScore + G-Eval
& 0.545 & 0.528 & 0.760
& 0.379 & 0.373 & 0.695
& 0.404 & 0.400 & \textbf{1.033} \\
\addlinespace[2pt]
\rowcolor{gray!10}
\ours
& \betterbg{\textbf{0.575}} & \underline{0.550} & \betterbg{\textbf{0.645}}
& \betterbg{\textbf{0.431}} & \betterbg{\underline{0.427}} & \betterbg{\textbf{0.569}}
& \betterbg{\underline{0.436}} & \betterbg{\underline{0.447}} & 1.075 \\
\rowcolor{gray!10}
\quad w/o self-merging
& \betterbg{\underline{0.572}} & 0.548 & \betterbg{\underline{0.653}}
& \betterbg{\underline{0.426}} & \betterbg{\textbf{0.430}} & \betterbg{\underline{0.640}}
& \betterbg{\textbf{0.440}} & \betterbg{\textbf{0.453}} & 1.045 \\
\midrule
\multicolumn{10}{c}{\textbf{Audex-2B}} \\
\midrule
Prompting (greedy)
& 0.249 & 0.239 & 2.169
& 0.208 & 0.225 & 1.047
& 0.374 & 0.377 & 1.286 \\
Prompting (avg@8)
& 0.305 & 0.305 & 1.936
& 0.206 & 0.219 & \underline{0.923}
& 0.431 & 0.463 & 1.079 \\
AQAScore
& 0.289 & 0.385 & 2.267
& \underline{0.275} & \textbf{0.294} & 2.332
& 0.345 & 0.459 & 2.557 \\
G-Eval
& 0.332 & 0.359 & \underline{1.819}
& 0.126 & 0.152 & 1.039
& 0.511 & 0.512 & \underline{0.967} \\
AQAScore + G-Eval
& 0.357 & 0.400 & 1.982
& 0.263 & 0.285 & 1.121
& 0.449 & 0.502 & 1.219 \\
\addlinespace[2pt]
\rowcolor{gray!10}
\ours
& \betterbg{\textbf{0.416}} & \betterbg{\textbf{0.417}} & \betterbg{\textbf{1.784}}
& \betterbg{\underline{0.275}} & 0.288 & \betterbg{\textbf{0.920}}
& \betterbg{\underline{0.550}} & \betterbg{\underline{0.566}} & 1.002 \\
\rowcolor{gray!10}
\quad w/o self-merging
& \betterbg{\underline{0.405}} & \betterbg{\underline{0.405}} & 1.870
& \betterbg{\textbf{0.278}} & \underline{0.290} & 1.054
& \betterbg{\textbf{0.566}} & \betterbg{\textbf{0.568}} & \betterbg{\textbf{0.950}} \\
\bottomrule
\end{tabularx}
\endgroup
\caption{Judging performance on audio-text relevance (Audio-REL), audio quality (Audio-OVL), and speech quality (Speech-OVL).
Formatting follows Table~\ref{tab:understanding}. Detailed results for each benchmark are provided in Appendix~\ref{app:full_judge_results}.}
\label{tab:judge-overall}
\end{table*}

\subsection{Main Results}
\label{sec:main_results}
We report the results in Tables~\ref{tab:understanding}-\ref{tab:tts-zero-shot}.

\paragraph{Understanding.}

As shown in Table~\ref{tab:understanding}, \ours achieves comparable or better performance than the base models on most understanding benchmarks.
The average accuracy across four benchmarks increases from 66.73 to 69.08 on UNISON and from 59.37 to 63.40 on Audex-2B.

Self-consistency improves performance on some benchmarks, but its benefits are inconsistent.
For example, self-consistency improves MMAU accuracy over greedy decoding on both models, but performs worse on MMAR and TREA.
This may be due to hallucinations introduced during reasoning.
In TREA, we observe cases where CoT counts repeated occurrences of the same sound as distinct events, leading to incorrect answers.
In contrast, \ours outperforms self-consistency in average accuracy on both UNISON (69.08 vs.\ 65.77) and Audex-2B (63.40 vs.\ 59.49), while requiring only a single inference instead of eight forwards.
Although self-consistency achieves higher accuracy on MMAU Test-mini and MMSU for Audex-2B, \ours still outperforms the single-pass baseline on all evaluated benchmarks and surpasses self-consistency on the remaining benchmarks.

The effect of model merging differs across architectures and benchmarks.
For UNISON, the unmerged model underperforms the base model on MMAU Test, MMAR, and TREA.
After merging, performance improves beyond the base model on MMAU Test and TREA, while MMAR recovers to a comparable level (63.00 vs.\ 63.10).
In contrast, the results on Audex-2B are mixed, with merging improving performance on some benchmarks but reducing it on others, while average accuracy remains similar (63.40 vs.\ 63.44).
Nevertheless, both the merged and unmerged variants outperform the base model and SC in average accuracy.

\paragraph{Judging.}

Table~\ref{tab:judge-overall} reports the average performance across benchmarks for each of the three tasks: Audio-OVL, Speech-OVL, and Audio-REL. 
Due to the page limit, detail results for each benchmark are provided in Appendix~\ref{app:full_judge_results}.

Direct prompting with greedy decoding generally performs the worst, while averaging eight sampled ratings (avg@8) improves performance but requires more inference compute. 
The AQAScore based method and the G-Eval based method have different strengths. 
For example, AQAScore achieves higher correlations on UNISON Audio-REL, while G-Eval performs better on Audex-2B Speech-OVL. 
We therefore combine their scores to construct pseudo labels for self-training, as described in \S\ref{subsec:method_understanding_judging}. 
Compared with these combined scores, \ours achieves better performance on most metrics and comparable results on the others. 
For example, on UNISON Audio-OVL, Pearson correlation increases from 0.379 to 0.431, Spearman correlation from 0.373 to 0.427, and MSE decreases from 0.695 to 0.569. 
On Audex-2B Speech-OVL, Pearson correlation also increases from 0.449 to 0.550. 
These results show that the improvements do not come from combining the two scoring methods alone. 
Self-training on their combined scores leads to further gains.

Model merging has a relatively small effect on judging performance.
Across both architectures, the merged and unmerged variants achieve comparable results, with each performing better on different benchmarks and metrics.
This suggests that the judging improvements are largely preserved regardless of whether model merging is applied.

\providecommand{\betterbg}[1]{\cellcolor{green!20}#1}
\begin{table*}[t]
\centering
\begingroup
\scriptsize
\setlength{\tabcolsep}{3pt}
\renewcommand{\arraystretch}{1.0}
\begin{adjustbox}{width=\textwidth}
\begin{tabular}{l*{7}{c}@{\hspace{8pt}}*{6}{c}}
\toprule
\textbf{Method}
& \multicolumn{7}{c}{\textbf{AudioCaps}}
& \multicolumn{6}{c}{\textbf{AudioTime}} \\
\cmidrule(lr){2-8} \cmidrule(l){9-14}
& \tiny FAD$\downarrow$ & \tiny FD$\downarrow$ & \tiny KL$\downarrow$ & \tiny IS$\uparrow$ & \tiny CLAP$\uparrow$ & \tiny \shortstack{CoTTAGE\\REL$\uparrow$} & \tiny \shortstack{CoTTAGE\\OVL$\uparrow$} & \tiny \shortstack{Order\\Error Rate$\downarrow$} & \tiny \shortstack{Duration\\L1 Error$\downarrow$} & \tiny \shortstack{Frequency\\L1 Error$\downarrow$} & \tiny \shortstack{Timestamp\\F1$\uparrow$} & \tiny \shortstack{CoTTAGE\\REL$\uparrow$} & \tiny \shortstack{CoTTAGE\\OVL$\uparrow$} \\
\midrule
\multicolumn{14}{c}{\textbf{UNISON}} \\
\midrule
Base & \textbf{1.56} & \textbf{15.51} & 1.445 & 11.57 & 0.496 & 3.940 & 3.229 & 0.754 & \underline{2.471} & 1.193 & 0.730 & 3.136 & 2.927 \\
\addlinespace[1pt]
\rowcolor{gray!10}
\ours & \underline{1.70} & \underline{15.85} & \betterbg{\underline{1.403}} & \betterbg{\textbf{11.78}} & \betterbg{\underline{0.513}} & \betterbg{\underline{4.056}} & \betterbg{\underline{3.272}} & \betterbg{\underline{0.690}} & \betterbg{\textbf{2.436}} & \betterbg{\underline{1.096}} & \betterbg{\underline{0.775}} & \betterbg{\underline{3.321}} & \betterbg{\underline{3.028}} \\
\rowcolor{gray!10}
\quad w/o self-merging & 2.21 & 16.49 & \betterbg{\textbf{1.381}} & \betterbg{\underline{11.76}} & \betterbg{\textbf{0.517}} & \betterbg{\textbf{4.086}} & \betterbg{\textbf{3.292}} & \betterbg{\textbf{0.634}} & 2.477 & \betterbg{\textbf{1.067}} & \betterbg{\textbf{0.785}} & \betterbg{\textbf{3.410}} & \betterbg{\textbf{3.109}} \\
\midrule
\multicolumn{14}{c}{\textbf{Audex-2B}} \\
\midrule
Base & 7.07 & \underline{41.75} & 2.254 & 5.54 & 0.281 & 3.062 & 2.748 & \underline{0.976} & \textbf{3.001} & \textbf{1.461} & \underline{0.476} & 2.322 & \underline{2.668} \\
\addlinespace[1pt]
\rowcolor{gray!10}
\ours & \betterbg{\underline{6.30}} & \betterbg{\textbf{39.81}} & \betterbg{\textbf{2.046}} & \betterbg{\textbf{5.78}} & \betterbg{\textbf{0.322}} & \betterbg{\textbf{3.250}} & \betterbg{\textbf{2.831}} & \betterbg{\textbf{0.966}} & \underline{3.074} & 1.541 & \betterbg{\textbf{0.491}} & \betterbg{\textbf{2.391}} & \betterbg{\textbf{2.708}} \\
\rowcolor{gray!10}
\quad w/o self-merging & \betterbg{\textbf{6.01}} & 41.84 & \betterbg{\underline{2.176}} & \betterbg{\underline{5.64}} & \betterbg{\underline{0.319}} & \betterbg{\underline{3.165}} & \betterbg{\underline{2.760}} & 0.984 & 3.337 & \underline{1.521} & 0.454 & \betterbg{\underline{2.384}} & 2.639 \\
\bottomrule
\end{tabular}
\end{adjustbox}
\endgroup
\caption{Performance on TTA benchmarks.
Formatting follows Table~\ref{tab:understanding}.}
\label{tab:audio-generation}
\end{table*}

\paragraph{Text-to-Audio.}

For TTA, our method improves performance on most AudioCaps metrics across both architectures, including CLAP, IS, KL, and CoTTAGE scores.
In particular, CLAP increases from 0.496 to 0.513 for UNISON and from 0.281 to 0.322 for Audex-2B. 
CoTTAGE scores also improve for both relevance and overall quality. 
For UNISON, FAD and FD increase slightly, from 1.56 to 1.70 and from 15.51 to 15.85, respectively. 
This pattern is also observed with prior TTA studies using external data ~\cite{tango2,t2a-feedback}, which show that gains in semantic alignment and perceptual quality can occur with degradation in distribution-level metrics.

On AudioTime, the results differ across architectures.
For UNISON,~\ours improves upon the base model across all six metrics.
In contrast, Audex-2B shows mixed results, with improvements in event ordering, timestamp F1, and CoTTAGE scores, but higher duration and frequency errors.
To better understand this limitation, we evaluate simple diagnostic baselines using the same evaluation pipeline.
We find that Audex-2B has a duration error of 3.00, higher than that of the silence baseline (2.54).
Its frequency error of 1.46 is also close to that of the silence baseline (1.50).
These observations suggest that Audex-2B has limited temporal generation capabilities, which may constrain self-improvement when the model relies on its own generated audio for supervision.

Model merging further affects how consistently these improvements are achieved.
For UNISON, the unmerged model slightly degrades AudioTime duration error compared with the base model (2.477 vs.\ 2.471), whereas merging reduces it to 2.436, resulting in improvements across all six AudioTime metrics.
Although the unmerged model performs better on event ordering, both variants improve upon the base model.
For Audex-2B, merging retains the improvements on AudioCaps while further improving CLAP and CoTTAGE scores over the unmerged variant.
It also improves event ordering and timestamp F1, which otherwise degrade without merging.
These results suggest that merging can help achieve more consistent improvements across evaluation metrics, even when the unmerged model performs better on some metrics.

\providecommand{\betterbg}[1]{\cellcolor{green!20}#1}

\begin{table}[t]
\centering
\begingroup
\scriptsize
\setlength{\tabcolsep}{2pt}
\renewcommand{\arraystretch}{1.05}
\begin{adjustbox}{width=\columnwidth}
\begin{tabular}{l*{2}{c}@{\hspace{4pt}}*{2}{c}@{\hspace{4pt}}*{2}{c}@{\hspace{4pt}}*{2}{c}}
\toprule
\textbf{Method}
& \multicolumn{2}{c}{\textbf{Seed-TTS}}
& \multicolumn{2}{c}{\textbf{CV3}}
& \multicolumn{2}{c}{\textbf{CV3-Hard}}
& \multicolumn{2}{c}{\textbf{LibriTTS}} \\
\cmidrule(lr){2-3} \cmidrule(lr){4-5} \cmidrule(lr){6-7} \cmidrule(l){8-9}
& \fontsize{5}{6}\selectfont WER$\downarrow$ & \fontsize{5}{6}\selectfont UTMOS$\uparrow$ & \fontsize{5}{6}\selectfont WER$\downarrow$ & \fontsize{5}{6}\selectfont UTMOS$\uparrow$ & \fontsize{5}{6}\selectfont WER$\downarrow$ & \fontsize{5}{6}\selectfont UTMOS$\uparrow$ & \fontsize{5}{6}\selectfont WER$\downarrow$ & \fontsize{5}{6}\selectfont UTMOS$\uparrow$ \\
\midrule
\multicolumn{9}{c}{\textbf{UNISON}} \\
\midrule
Base & 1.49 & \underline{3.33} & 3.26 & \underline{3.23} & 20.32 & \underline{2.74} & 5.46 & \underline{3.12} \\
\addlinespace[2pt]
\rowcolor{gray!10}
\ours & \betterbg{\textbf{1.46}} & \betterbg{\textbf{3.35}} & \betterbg{\textbf{3.11}} & \betterbg{\textbf{3.26}} & \betterbg{\underline{18.39}} & \betterbg{\textbf{2.80}} & \betterbg{\textbf{4.95}} & \betterbg{\textbf{3.16}} \\
\rowcolor{gray!10}
\quad w/o self-merging & \betterbg{\underline{1.47}} & \betterbg{\underline{3.33}} & \betterbg{\underline{3.12}} & 3.22 & \betterbg{\textbf{17.65}} & 2.71 & \betterbg{\underline{5.02}} & \betterbg{\underline{3.12}} \\
\midrule
\multicolumn{9}{c}{\textbf{Audex-2B}} \\
\midrule
Base & \underline{2.18} & \textbf{4.21} & 4.22 & 4.23 & \underline{6.03} & \underline{4.28} & 4.19 & \underline{4.13} \\
\addlinespace[2pt]
\rowcolor{gray!10}
\ours & \betterbg{\textbf{1.96}} & \betterbg{\textbf{4.21}} & \betterbg{\underline{3.75}} & \betterbg{\textbf{4.25}} & \betterbg{\textbf{5.79}} & \betterbg{\textbf{4.29}} & \betterbg{\underline{3.67}} & \betterbg{\textbf{4.14}} \\
\rowcolor{gray!10}
\quad w/o self-merging & 2.44 & 4.20 & \betterbg{\textbf{3.64}} & \betterbg{\textbf{4.25}} & 6.24 & 4.27 & \betterbg{\textbf{3.60}} & 4.12 \\
\bottomrule
\end{tabular}
\end{adjustbox}
\endgroup
\caption{Performance on conventional TTS benchmarks.
Formatting follows Table~\ref{tab:understanding}.}
\label{tab:tts-conventional}
\end{table}
\providecommand{\betterbg}[1]{\cellcolor{green!20}#1}
\begin{table}[t]
\centering
\begingroup
\scriptsize
\setlength{\tabcolsep}{3pt}
\renewcommand{\arraystretch}{1.05}
\begin{adjustbox}{width=\columnwidth}
\begin{tabular}{l*{3}{c}@{\hspace{8pt}}*{3}{c}@{\hspace{8pt}}*{3}{c}}
\toprule
\textbf{Method}
& \multicolumn{3}{c}{\textbf{Seed-TTS}}
& \multicolumn{3}{c}{\textbf{CV3}}
& \multicolumn{3}{c}{\textbf{CV3-Hard}} \\
\cmidrule(lr){2-4} \cmidrule(lr){5-7} \cmidrule(l){8-10}
& \tiny WER$\downarrow$ & \tiny UTMOS$\uparrow$ & \tiny SIM$\uparrow$ & \tiny WER$\downarrow$ & \tiny UTMOS$\uparrow$ & \tiny SIM$\uparrow$ & \tiny WER$\downarrow$ & \tiny UTMOS$\uparrow$ & \tiny SIM$\uparrow$ \\
\midrule
\multicolumn{10}{c}{\textbf{UNISON}} \\
\midrule
Base & \underline{2.27} & 3.12 & \underline{51.51} & 5.42 & 2.97 & \textbf{41.34} & 36.00 & 2.49 & \textbf{38.14} \\
\addlinespace[2pt]
\rowcolor{gray!10}
\ours & \betterbg{\textbf{2.25}} & \betterbg{\textbf{3.15}} & \betterbg{\textbf{51.69}} & \betterbg{\underline{5.14}} & \betterbg{\underline{2.99}} & \underline{40.77} & \betterbg{\textbf{32.06}} & \betterbg{\textbf{2.54}} & \underline{37.67} \\
\rowcolor{gray!10}
\quad w/o self-merging & 2.64 & \betterbg{\underline{3.14}} & 50.41 & \betterbg{\textbf{4.92}} & \betterbg{\textbf{3.00}} & 39.27 & \betterbg{\underline{32.27}} & \betterbg{\underline{2.52}} & 36.45 \\
\bottomrule
\end{tabular}
\end{adjustbox}
\endgroup
\caption{Performance on zero-shot TTS benchmarks.
Formatting follows Table~\ref{tab:understanding}.}
\label{tab:tts-zero-shot}
\end{table}

\paragraph{Text-to-Speech.}
We evaluate both UNISON and Audex-2B on conventional TTS. Since Audex-2B lacks support for voice cloning, we evaluate zero-shot TTS only on UNISON.

On conventional TTS (Table~\ref{tab:tts-conventional}), \ours reduces WER across all four benchmarks on both models, while maintaining comparable or slightly higher UTMOS scores.
For example, WER on CV3-Hard decreases from 20.32\% to 18.39\% for UNISON and from 6.03\% to 5.79\% for Audex-2B. 
Although the unmerged models achieve lower WER on some benchmarks, self-merging leads to more consistent improvements. 
In particular, without merging, Audex-2B shows higher WER than the base model on Seed-TTS and CV3-Hard. 
The merged model achieves lower WER than the base model on both benchmarks.

On zero-shot TTS (Table~\ref{tab:tts-zero-shot}), the merged UNISON achieves lower WER and slightly higher UTMOS scores across all three benchmarks, despite our post-training not explicitly targeting voice cloning. 
The improvement is most noticeable on CV3-Hard, where WER decreases from 36.00\% to 32.06\%. 
Without merging, WER increases on Seed-TTS and speaker similarity decreases across all three benchmarks. 
In contrast, self-merging maintains speaker similarity scores close to those of the base model while improving WER across the evaluated benchmarks.

\begin{figure}
    \centering
    \includegraphics[width=\linewidth]{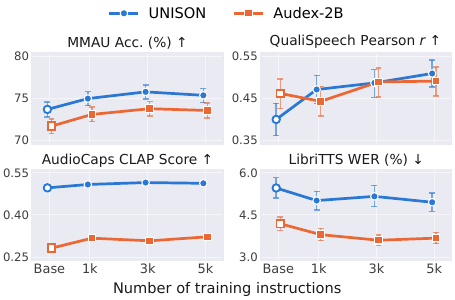}
    \caption{Data scaling analysis on representative benchmarks. 
    Error bars indicate 95\% confidence
intervals.}
    \label{fig:data_scaling}
\end{figure}

\subsection{Data Scaling}
We investigate how the number of self-generated instructions affects model performance. 
Since these instructions are used to construct QA examples, judging supervision, and preference pairs, we vary their number from 1K to 5K.
Figure 3 shows the results on four representative benchmarks covering understanding, judging, audio generation, and speech generation.
Error bars indicate 95\% confidence intervals computed from 2,000 bootstrap resamples of test items.

For both models, using 1K self-generated instructions already improves performance over the base models on most benchmarks.
Increasing the number of instructions to 5K brings further gains on several benchmarks, although the magnitude varies across tasks.
These results show that~\ours can effectively improve multiple capabilities with only a small amount of self-generated data.

\section{Conclusion}
In this work, we introduce a self-improving post-training framework that leverages the understanding, judging, and generation capabilities of unified audio models to improve themselves without external data or teacher models. 
Experiments across multiple benchmarks show that our framework generally improves understanding, judging, and generation.
We further find that self-merging leads to more consistent improvements across benchmarks.
Although fine-tuning without merging can achieve better results on individual tasks, it may also degrade performance below the base model on others. 
In contrast, self-merging helps reduce such regressions, providing a more reliable strategy for achieving broad improvements through self-improvement.
Overall, our findings highlight the potential of leveraging existing model capabilities for self-improvement without external supervision.

\section*{Limitations}
We conduct all experiments in English, as most of the benchmarks we evaluate on are in English. 
This setting is also constrained by the base models themselves.
Because our framework relies entirely on self-generated instructions and audio, it can only operate in languages that the base model can already understand and generate reliably, and the models we study do not necessarily support high-quality speech or audio generation in other languages.
Extending the framework beyond English therefore depends on the multilingual ability of the base model. 
Future work can examine whether these gains transfer to other languages.

Because \ours uses no external data or teacher model, it can only refine capabilities that the base model already possesses to some degree.
When the base model has limited control over a target property, self-improvement may yield little or no improvement on that property.
For example, on AudioTime, \ours does not improve Audex-2B’s control over event duration or repetition count. 
The capabilities of each base model also constrain our evaluation.
We evaluate zero-shot voice cloning only on UNISON, because Audex-2B does not support this task.
These results indicate that self-improvement relies on the capabilities of the base model.

In our experiments, one round of self-improving post-training yields gains in understanding, judging, and generation.
We leave for future work the question of how to build on these gains over multiple rounds, with the updated model generating new training data for each subsequent round.

\section*{Ethical Considerations}

Our work focuses on improving the general capabilities of unified audio models through self-improving post-training. 
While we evaluate zero-shot voice cloning as an existing capability of the base model, our method does not explicitly optimize speaker imitation. 
Nevertheless, such models could be misused to generate speech that impersonates individuals without their consent. 
We therefore emphasize that voice cloning should only be used with appropriate authorization and that safeguards against misuse are important for real-world deployment.

\section*{Acknowledgments}
This work was supported by the Ministry of Education (MOE) of Taiwan under the project Taiwan Centers of Excellence in Artificial Intelligence, through the NTU Artificial Intelligence Center of Research Excellence (NTU AI-CoRE).
 
We thank the National Center for High-performance Computing (NCHC) of the National Applied Research Laboratories (NARLabs) in Taiwan for providing computational and storage resources.


\bibliography{custom}

\appendix

\section{Evaluation Benchmarks}
\label{app:benchmarks}

\subsection{Audio Understanding}

\paragraph{MMAU.}
MMAU is an audio understanding and reasoning benchmark with 10,000 questions, split into a 9,000-question test set and a 1,000-question test-mini set~\cite{mmau}.
The questions span speech, environmental sounds, and music, and assess 27 skills in information retrieval and reasoning.

\paragraph{MMAR.}
MMAR contains 1,000 questions designed to assess multi-step reasoning over speech, sounds, music, and their mixtures~\cite{mmar}.
Audio is collected from online videos, and questions are organized into four reasoning layers: signal, perception, semantics, and culture.

\paragraph{MMSU.}
MMSU evaluates spoken language understanding through 5,000 questions spanning 47 tasks, comprising 24 perception tasks and 23 reasoning tasks~\cite{mmsu}.
Its design draws on linguistic theory and covers semantic and phonological information, speaker characteristics, and speaking styles.
Tasks require models to use both linguistic content and acoustic cues, including prosody, emotion, and speaking speed.

\paragraph{TREA.}
TREA evaluates temporal reasoning over environmental sounds~\cite{trea}.
It contains 600 questions, equally divided among duration comparison, event ordering, and event counting.
The questions ask which sound lasted longest, which sound came next, or how many different sounds were in the clip. 
The audio is constructed by combining recordings from ESC-50~\cite{esc}.

\subsection{Audio Judge}
\paragraph{BVCC.}
BVCC evaluates the naturalness of English speech in the main track of the VoiceMOS Challenge 2022~\cite{bvcc,voicemos2022}.
It includes natural speech and samples from text-to-speech and voice conversion systems.
Eight listeners rated each utterance on a five-point scale, and their ratings were averaged to obtain a mean opinion score (MOS).
The test split contains 1,066 utterances.

\paragraph{QualiSpeech.}
QualiSpeech evaluates the quality of English speech~\cite{qualispeech}.
It includes natural recordings from GigaSpeech~\cite{gigaspeech} and NISQA~\cite{nisqa}, recordings with simulated degradations, and synthetic speech from BVCC and more recent text-to-speech systems.
The dataset provides five-point ratings for noise, distortion, speed, continuity, listening effort, naturalness, and overall quality, along with human-reviewed descriptions of speech quality.
The test split contains 1,852 utterances.

\paragraph{PAM-Audio.}
PAM-Audio evaluates overall audio quality and audio-text relevance~\cite{pam}.
It uses 100 captions from the AudioCaps~\cite{audiocaps} test split, each paired with the original recording and four generated clips.
Listeners rated each clip for overall quality and relevance to its caption on two five-point scales.
The evaluation set contains 500 clips.

\paragraph{RELATE.}
RELATE evaluates audio-text relevance~\cite{relate}.
It includes natural AudioCaps recordings and clips synthesized from AudioCaps captions.
Listeners rated how well each clip matched its caption on an 11-point scale from 0 to 10.
In our experiments, we linearly rescale the human relevance scores for RELATE and XACLE to range from 1 to 5.
The dataset also provides separate ratings for whether the described sounds are present and whether they occur in the described order.
The AudioCaps test split contributes 2,598 pairs with overall relevance ratings, of which 1,311 form the RELATE benchmark test split.

\paragraph{XACLE.}
XACLE evaluates audio--text relevance~\cite{xacle}.
It includes natural recordings from AudioCaps and YouTube, along with clips generated from their captions.
Listeners rated how well each clip matched its English caption on an 11-point scale from 0 to 10.
In our experiments, we linearly rescale the human relevance scores for RELATE and XACLE to range from 1 to 5.
The test split contains 3,000 utterances.

\paragraph{Clotho.}
The Clotho test set evaluates overall audio quality and audio-text relevance~\cite{clotho,clotho-mos}.
It uses 100 captions from Clotho, each paired with the original recording and four generated clips.
Listeners rated each clip for overall quality and relevance to its caption on five-point scales.
The released evaluation set contains 500 clips.

\subsection{Text-to-Audio}

\paragraph{AudioCaps.}
AudioCaps was introduced for audio captioning and pairs audio clips from AudioSet~\cite{audioset} with descriptions written by human annotators~\cite{audiocaps}.
The recordings are approximately 10 seconds long and cover sounds from real-world YouTube videos.
Annotation guidelines emphasize audible content and discourage visual-only information and speech transcription.
The captions provide natural language descriptions of sound events and their acoustic context.

\paragraph{AudioTime.}
AudioTime is a temporally aligned audio-text dataset for evaluating temporal control in TTA generation~\cite{audiotime}.
It covers event ordering, duration, occurrence frequency, and timestamps.
Each category contains 5,000 training examples and 500 test examples, with audio clips of up to 10 seconds, temporal metadata, and corresponding captions.

\subsection{Text-to-Speech}

\paragraph{SEED-TTS-Eval.}
SEED-TTS-Eval is a benchmark for zero-shot speech synthesis that includes English and Mandarin speech~\cite{seed-tts}.
Each example pairs a reference utterance with a target utterance from the same speaker. The model uses the reference speech and target text to synthesize the target utterance.
In this work, we evaluate on the English subset (\texttt{test-en}).

\paragraph{CV3-Eval.}
CV3-Eval is a benchmark for zero-shot speech synthesis under real-world conditions~\cite{cosyvoice3}.
It covers multilingual voice cloning, cross-lingual voice cloning, and emotion cloning.
The standard multilingual evaluation includes 500 examples per language across nine languages.
The benchmark also includes hard subsets for Chinese and English, with target texts containing rare words, tongue twisters, and domain-specific terms.
In this work, we evaluate on the standard English subset (\texttt{en}) and the challenging English subset (\texttt{hard-en}).

\paragraph{LibriTTS.}
LibriTTS is a multi-speaker English corpus designed for text-to-speech research~\cite{libritts}.
It contains approximately 585 hours of read speech from audiobooks with corresponding transcripts.
The corpus provides two test splits, \texttt{test-clean} and \texttt{test-other}.
In this work, we evaluate on \texttt{test-clean}.

\section{Implementation Details}
\label{app:implementation_details}
\subsection{Setups}

We conduct experiments on two unified audio models with different architectures: UNISON~\cite{unison} and Audex-2B~\cite{audex}. UNISON conditions its MM-DiT blocks on the Qwen2.5-Omni-7B Thinker~\cite{qwen25omni}, while Audex-2B uses a shared LLM backbone with modality-specific encoders and decoders.

For the main experiments, we generate 5,000 instructions, which are then used to construct both generation and understanding data.
Each pipeline uses equal proportions of audio instructions and speech examples, with the latter including both utterances and dialogues.
To construct multi-speaker dialogues, we synthesize each turn separately and concatenate the resulting waveforms.
For UNISON, we construct a bank of synthetic reference voices and
screen out acoustically similar candidates using MFCC-based DTW
distances computed over the same reference sentence, drawing on
text-dependent speaker verification methods~\citep{dsp_speaker_verification}.
We assign each dialogue speaker a distinct reference clip and
condition all turns by that speaker on the same clip.
Since Audex-2B does not support reference-conditioned voice cloning, we instead assign each speaker a distinct signal-processing preset and apply it to the synthesized speech.
These presets adjust formants, pitch, pitch range, and duration, with the same preset reused across all turns by a given speaker.

We use LoRA~\citep{lora} for LLM-based fine-tuning. For generation training, Audex-2B uses autoregressive token-level DPO~\citep{dpo}, while UNISON uses flow-matching DPO~\citep{diffusion-dpo}. 
The two formulations define DPO $\beta$ differently. 
We use interval guidance~\citep{interval_guidance}  when sampling training data with UNISON.
During this sampling process, we apply classifier-free guidance at normalized noise levels between 0.2 and 0.9 and use the conditional prediction alone outside this range.

For Audex-2B, we fine-tune the understanding and generation branches separately, starting from the same base model. We then add both updates to the base weights, scaling each by 0.5.
For UNISON, we average the fine-tuned LALM weights with the base LALM weights before training the DiT. 
Following UNISON's original setup~\citep{unison}, we freeze the merged LALM during DiT training and maintain an exponential moving average (EMA)~\citep{ema} of the DiT weights. 
After training, we average the saved EMA weights with the base DiT weights.

All experiments run on a single NVIDIA H200 GPU. Tables~\ref{tab:sampling_hyperparameters} and~\ref{tab:training_hyperparameters} summarize the data sampling and training settings.

\begin{table*}[t]
\centering
\small
\resizebox{0.8\textwidth}{!}{%
\begin{tabular}{lcc}
\toprule
Parameter & Audex-2B & UNISON \\
\midrule
\multicolumn{3}{l}{\textit{Text generation}} \\
Temperature & 0.8 & 0.8 \\
Top-$p$ & 0.9 & 0.95 \\
Semantic variants per instruction & 3 & 3 \\
\midrule
\multicolumn{3}{l}{\textit{Audio generation}} \\
Candidates per instruction & 8 & 8 \\
Temperature & 1.0 & - \\
Top-$k$ &  80 & - \\
CFG scale & Audio: 3.0; Speech: 1.0 & 6.0 \\
Inference steps & - & 50 \\
Interval guidance & - &  $\sigma\in[0.2,0.9]$ \\
\midrule
\multicolumn{3}{l}{\textit{QA and preference pair construction}} \\
QA source clips per instruction & 1 & 1 \\
MCQs generated per audio & 8  & 8  \\
Votes per MCQ & 8 & 8 \\
Minimum consensus votes & 4 & 4 \\
Preference Pair Score Margin (1–5) & 0.25 & 0.25 \\
\bottomrule
\end{tabular}%
}
\caption{Sampling and data-selection settings. }
\label{tab:sampling_hyperparameters}
\end{table*}
\begin{table*}[t]
\centering
\begin{tabular}{lcc}
\toprule
Parameter & Audex-2B & UNISON \\
\midrule
Stage 2 learning rate & $5\times10^{-5}$ & $1\times10^{-4}$ \\
Stage 2 epochs & 1 & 1 \\
Stage 2 batch size & 32 & 32 \\
Stage 2 LoRA rank / alpha & 64 / 128 & 64 / 128 \\
\midrule
Stage 3 learning rate & $1\times10^{-5}$ & $1\times10^{-5}$ \\
Stage 3 epochs & 1 & 1 \\
Stage 3 batch size & 64 & 64 \\
Stage 3 LoRA rank / alpha & 64 / 128 & Full DiT fine-tuning \\
Stage 3 DPO $\beta$ & 0.1 & 2,000 \\
\midrule
Optimizer / weight decay & AdamW / 0.01 & AdamW / 0.01 \\
LR warmup & 5\% & 5\% \\
Merging weights (Stage 2 / Stage 3) & 0.5 / 0.5 & 0.5 / 0.5 \\
\bottomrule
\end{tabular}
\caption{Training hyperparameters.}
\label{tab:training_hyperparameters}
\end{table*}

\subsection{Prompts}

\begin{table*}[t]
    \centering

    \footnotesize
    \setlength{\tabcolsep}{7pt}
    \renewcommand{\arraystretch}{1.25}

    \begin{tabularx}{0.95\textwidth}{
        >{\raggedright\arraybackslash\bfseries}p{0.18\textwidth}
        >{\raggedright\arraybackslash}X
    }
        \toprule
        Category & \textbf{Prompt} \\
        \midrule

        General Audio &
        Generate \texttt{\{k\}} sound event descriptions suitable for text-to-audio generation. Each description should describe clearly audible sound events that fit within a 10-second clip. Cover a wide range of settings: wildlife, weather, environmental sound, machines, musical instruments, human activity, and so on. Reply with ONLY a JSON list of \texttt{\{k\}} strings. \\
        \midrule

        Temporal Event List &
        Generate \texttt{\{k\}} sound scenes for text-to-audio generation. Describe each scene as two to four distinct sound events separated by ' ; ', in the order they should be heard. Keep every event a short lowercase sound description of at most eight words. Do NOT write any numbers, timings, durations or counts, only the sounds themselves. Reply with ONLY a JSON list of \texttt{\{k\}} strings. \\
        \midrule

        Overlap and Simultaneity &
        Generate \texttt{\{k\}} sound scene descriptions for text-to-audio generation. Each scene should contain two sounds that overlap in time. Make it clear that the two sounds happen at the same time, such as a longer sound continuing while a shorter sound occurs over it. Reply with ONLY a JSON list of \texttt{\{k\}} strings. \\
        \midrule

        Negation and Absence &
        Generate \texttt{\{k\}} sound scene descriptions for text-to-audio generation. Each description should include one or two sounds that are present and exactly one related sound that is absent. State the absent sound using 'no', 'without', or 'never'. Choose a sound that would normally be expected to occur with the other sounds. Reply with ONLY a JSON list of \texttt{\{k\}} strings. \\
        \midrule

        Multiple Sound Sources &
        Generate \texttt{\{k\}} sound scene descriptions for text-to-audio generation. Each scene should include three to five different sound sources, and all of them should be clearly audible in the clip. You may state the exact number of sources when useful. Keep the sounds distinct enough to tell them apart. Reply with ONLY a JSON list of \texttt{\{k\}} strings. \\
        \bottomrule
    \end{tabularx}

\caption{Prompts for generating training audio across different categories. The placeholder \texttt{\{k\}} denotes the number of requested descriptions. For the temporal event list, we randomly assign each event a timestamp, duration, or occurrence count and add it to the caption with a fixed template.}
    \label{tab:training_audio_prompts}
\end{table*}

\begin{table*}[t]
    \centering

    \footnotesize
    \setlength{\tabcolsep}{7pt}
    \renewcommand{\arraystretch}{1.25}

    \begin{tabularx}{0.95\textwidth}{
        >{\raggedright\arraybackslash\bfseries}p{0.18\textwidth}
        >{\raggedright\arraybackslash}X
    }
        \toprule
        Category & \textbf{Prompt} \\
        \midrule

        General Speech &
        Generate \texttt{\{k\}} English sentences suitable for text-to-speech generation. Each sentence should be natural and speakable within a 10-second clip. Reply with ONLY a JSON list of \texttt{\{k\}} strings. \\
        \midrule

        Lexical Ambiguity and Idioms &
        Generate \texttt{\{k\}} English sentences for text-to-speech generation. Each sentence should contain exactly one meaningful lexical challenge: a common idiom, a polysemous word used in one specific sense, a homophone whose intended spelling is clear from context, or a short wordplay expression. Provide enough natural context inside the sentence to make the intended interpretation unambiguous, but do not explain or define the expression. Avoid obscure cultural references and quotations. Keep each item to one self-contained sentence that can be spoken within 10 seconds. Reply with ONLY a JSON list of \texttt{\{k\}} strings. \\
        \midrule

        Contrastive Phonetic Content &
        Generate \texttt{\{k\}} natural English sentences for text-to-speech generation. Each sentence should contain one clearly identifiable set of two or three easily confused words that must be pronounced distinctly. Use minimal or near-minimal contrasts such as a single vowel difference, a single consonant difference, a voicing contrast, or a confusable number pair. Include every contrasting word explicitly in the same sentence and provide enough context to make each word appropriate. Do not use exact homophones, random word lists, phonetic notation, or tongue twisters. Keep each item to one sentence that can be spoken within 10 seconds. Reply with ONLY a JSON list of \texttt{\{k\}} strings. \\
        \bottomrule
    \end{tabularx}
    \caption{Prompts for generating training speech across different categories. The placeholder \texttt{\{k\}} denotes the number of requested sentences.}
    \label{tab:training_speech_prompts}
\end{table*}

\begin{table*}[t]
    \centering

    \footnotesize
    \setlength{\tabcolsep}{7pt}
    \renewcommand{\arraystretch}{1.2}

    \begin{tabularx}{0.98\textwidth}{
        >{\raggedright\arraybackslash\bfseries}p{0.18\textwidth}
        >{\raggedright\arraybackslash}X
    }
        \toprule
        Category & \textbf{Prompt} \\
        \midrule

        Dialogue Planning &
        Write exactly \texttt{\{k\}} substantially different outlines for multi-speaker dialogues.

        \par\smallskip
        Each dialogue should be designed to naturally exercise the target capability specified below. Use the capability key as the task category, and use the capability definition to understand what kind of information, reasoning, or interaction the future dialogue should require.

        \par\smallskip
        Capability key: \texttt{\{capability\}}

        \par
        Capability definition: \texttt{\{capability\_description\}}

        \par\smallskip
        At this stage, generate planning briefs only. Do not write any part of the actual conversation.

        \par\smallskip
        \textbf{Requirements:}
        \par
        -- Each brief should describe what the future conversation is about, the situation in which it takes place, how it develops, and what the final outcome it reaches.
        \par
        -- Choose a concrete main topic and a plausible setting.
        \par
        -- Specify 2 to 6 participant roles.
        \par
        -- Write the summary in third-person planning prose. It should cover the premise, the important information exchanged, the progression of the situation, and the final outcome.
        \par
        -- Design the situation so that demonstrating the target capability is necessary or useful for successfully completing the conversation.
        \par
        -- Make all \texttt{\{k\}} briefs different from one another.
        \par
        -- The topic is not limited to everyday, professional, civic, educational, travel, service, hobby, or family contexts, and may extend beyond generic meetings, cafes, appointments, project deadlines, or reservations.
        \par
        -- The entire situation must be expressible through spoken dialogue alone. Do not rely on narration, visual information, or background sounds to make the scenario understandable.

        \par\smallskip
        Reply with ONLY a JSON list of planning briefs. Each item must contain exactly these five fields and no additional fields, especially no dialogue or script field:

        \par

        \texttt{[\{"topic": "...", "setting": "...", "participants": ["...", "..."], "summary": "...", "key\_details": ["...", "..."]\}]}

        \\

        \midrule

        Dialogue Generation &
        Write exactly one English multi-speaker dialogue based on the planning context below.

        \par\smallskip
        Capability key: \texttt{\{capability\}}

        \par
        Capability definition: \texttt{\{capability\_description\}}

        \par\smallskip
        Planning context (follow it closely; do not replace it with a generic scenario):

        \par
        \texttt{\{planning\_context\}}

        \par\smallskip
        \textbf{Requirements:}
        \par
        -- Use 2 to 6 speakers and 5 to 10 turns.
        \par
        -- Adjacent turns must use different speakers.
        \par
        -- Every speaker must speak at least once.
        \par
        -- Include explicit spoken evidence needed for the requested capability.
        \par
        -- Construct the dialogue so the spoken turns contain all evidence needed to test the capability definition.
        \par
        -- Use natural facts, references, corrections, requests, or role clues.
        \par
        -- Do not include text that should not be spoken aloud.

        \par\smallskip
        Reply with ONLY a JSON list containing exactly one item. It must have this shape:

        \par

        \texttt{[\{"scenario": "...", "speakers": ["S1", "S2"], "turns": [\{"speaker": "S1", "text": "..."\}, \{"speaker": "S2", "text": "..."\}]\}]}

        \\

        \midrule

        Capability Definitions &
        \texttt{conversational\_fact\_retrieval}:
        Include explicit facts such as names, locations, quantities, plans, or preferences that can be retrieved from what the speakers say.

        \par\smallskip
        \texttt{role\_and\_relationship}:
        Include spoken clues that establish roles or relationships, such as customer and employee, teacher and student, coworkers, or relatives.

        \par\smallskip
        \texttt{deixis\_and\_coreference}:
        Include pronouns or referring expressions such as it, they, that one, there, or the earlier request, whose referent can be resolved from previous spoken turns.

        \par\smallskip
        \texttt{intent\_and\_speech\_act}:
        Include a clearly spoken request, refusal, promise, apology, warning, correction, suggestion, or confirmation whose communicative intent can be identified from the dialogue.

        \par\smallskip
        \texttt{dialogue\_scene\_reasoning}:
        Include spoken clues from which the likely setting or ongoing activity can be inferred. Do not rely on background sounds or narration. \\

        \bottomrule
    \end{tabularx}
    \caption{Prompts for generating multi-speaker training dialogues.
    Dialogue generation consists of two stages: dialogue planning and script generation.
    The placeholders \texttt{\{capability\}} and \texttt{\{capability\_description\}} are replaced with the corresponding capability key and definition listed in the last row of the table.
    The placeholder \texttt{\{planning\_context\}} is replaced with the planning brief generated in the first stage.}
    \label{tab:dialogue_generation_prompts}
\end{table*}

\begin{table*}[t]
    \centering

    \footnotesize
    \setlength{\tabcolsep}{6pt}
    \renewcommand{\arraystretch}{1.15}

    \begin{tabularx}{0.95\textwidth}{
        >{\raggedright\arraybackslash\bfseries}p{0.18\textwidth}
        >{\raggedright\arraybackslash}X
    }
        \toprule
        Category & \textbf{Prompt} \\
        \midrule

        Audio Question Generation &
        You are given an audio clip and its reference caption.

        \par
        Reference caption: \texttt{\{ref\}}

        \par\smallskip
        Listen carefully to the audio and generate \texttt{\{n\}} diverse multiple-choice questions that assess understanding of this specific audio clip. Use the caption to identify candidate facts, but confirm them from the audio. Do not ask about any detail that is not audibly verifiable. The trainee will receive the audio and question, but not the caption.

        \par\smallskip
        \textbf{Requirements:}
        \par
        -- Each question must require listening to the audio to answer correctly.
        \par
        -- Cover a variety of acoustic aspects, such as sound events, temporal order, the overall scene and ambience, speech (e.g., speaker characteristics, emotion, or spoken content), and music (e.g., instruments, tempo, rhythm, or mood) when present.
        \par
        -- Provide exactly four answer options for each question, with exactly one correct answer.
        \par
        -- Incorrect options should be plausible for a similar audio clip but clearly incorrect for this one. Good distractors often differ from the correct answer in just one detail, such as the order of two events, a count, a duration, or an attribute of a sound.

        \par\smallskip
        Reply with ONLY a JSON list in the following format:

        \par
        \texttt{[\{"question": "...", "options": ["...", "...", "...", "..."], "answer\_index": 0\}]} \\

        \midrule

        Dialogue Question Generation &
        You are given a multi-speaker dialogue and its reference transcript.

        \par
        Reference transcript: \texttt{\{ref\}}

        \par\smallskip
        Listen to the complete audio and generate \texttt{\{n\}} multiple-choice questions for the assigned capability below.

        \par
        Capability key: \texttt{\{capability\}}

        \par
        Capability definition: \texttt{\{capability\_description\}}

        \par\smallskip
        Use the transcript only to identify candidate facts. You should design the question, confirm each fact, and the correct answer from the audio.

        \par\smallskip
        \textbf{Requirements:}
        \par
        -- Every question must require listening to this dialogue.
        \par
        -- Do not ask about any detail that is not audibly verifiable.
        \par
        -- Do not generate speaker-count, turn-count, total-duration, timestamp, or utterance-start-time questions.
        \par
        -- Never refer to hidden labels such as S1, S2, Speaker A, or Speaker B.
        \par
        -- Refer to a voice by audible order or spoken evidence, such as 'the first speaker' or 'the person who mentioned the reservation'.
        \par
        -- Give exactly four distinct plausible options and exactly one correct answer.

        \par\smallskip
        Reply with ONLY a JSON list:

        \par
        \texttt{[\{"question": "...", "options": ["...", "...", "...", "..."], "answer\_index": 0\}]} \\

        \midrule

        Answer Verification &
        Listen to the audio. First describe briefly what you hear (one or two sentences). Then answer the question.

        \par\smallskip
        Question: \texttt{\{question\}}

        (A) \texttt{\{option A\}}
        (B) \texttt{\{option B\}}
        (C) \texttt{\{option C\}}
        (D) \texttt{\{option D\}}

        \par\smallskip
        End your reply with exactly: Final answer: (X) \\

        \midrule

        Programmatic Dialogue Questions &
        How many distinct speakers are heard in the dialogue?

        \par\smallskip
        How many speaking turns are heard in the dialogue?

        \par\smallskip
        Approximately how long is the complete dialogue?

        \par\smallskip
        Approximately when does the \texttt{\{ordinal\}} utterance begin? \\

        \midrule

        Multiple-Choice Understanding &
        Listen to the audio and answer the following question.

        \par\smallskip
        Question: \texttt{\{question\}}

        \par
        (A) \texttt{\{option A\}}
        (B) \texttt{\{option B\}}
        (C) \texttt{\{option C\}}
        (D) \texttt{\{option D\}}

        \par\smallskip
        Answer: \\

        \bottomrule
    \end{tabularx}

    \caption{Prompts used for training data construction, including question generation, answer verification, multiple-choice understanding.
    The placeholders \texttt{\{ref\}}, \texttt{\{question\}}, \texttt{\{instruction\}}, and \texttt{\{option A\}}--\texttt{\{option D\}} are replaced with the corresponding reference text, question, instruction, and answer options. The placeholders \texttt{\{k\}} and \texttt{\{n\}} denote the requested numbers of generated items.}
    \label{tab:training_data_prompts}

\end{table*}


\begin{table*}[t]
    \centering
    \footnotesize
    \setlength{\tabcolsep}{7pt}
    \renewcommand{\arraystretch}{1.2}

    \begin{tabularx}{0.95\textwidth}{
        >{\raggedright\arraybackslash\bfseries}p{0.18\textwidth}
        >{\raggedright\arraybackslash}X
    }
        \toprule
        Category & \textbf{Prompt} \\
        \midrule

        Audio Negative Instruction Generation &
        You will be given a description of a 10-second sound clip. Rewrite it into \texttt{\{k\}} DIFFERENT variants. Each variant should describe a clip that would not match the original description. For each variant, make one of the following changes while keeping everything else the same: change the order of events, change a timestamp or duration, change how many times a sound occurs, switch an attribute between two sounds, change a sound that is absent to present or a present sound to absent, add a new sound event, or remove one of the described sound events. Each change should be as small as possible, so the resulting clip remains highly similar to the original while still sounding clearly different. Make sure the variants are also different from one another.

        \par\smallskip
        Description: \texttt{\{instruction\}}

        \par\smallskip
        Reply with ONLY a JSON list of \texttt{\{k\}} strings. \\

        \midrule

        Speech Negative Instruction Generation &
        You will be given a sentence that a voice will speak aloud in a 10-second clip. Rewrite it into \texttt{\{k\}} DIFFERENT variants that sound very similar when spoken, but are not exactly the same words. A recording of the original sentence should not match any of the variants exactly. For each variant, make one of the following changes while leaving the rest of the sentence unchanged: replace a word with a word that sounds very similar when spoken (for example, 'seventy' can become 'seventeen', 'bell' can become 'bill', or 'lock' can become 'log'), or change how many times a repeated word appears. The replacement word must sound very similar to the original word when spoken. Do not use synonyms or words that sound noticeably different. Keep the length and rhythm as close to the original as possible, and make sure every variant sounds natural when spoken.

        \par\smallskip
        Sentence: \texttt{\{instruction\}}

        \par\smallskip
        Reply with ONLY a JSON list of \texttt{\{k\}} strings. \\

        \bottomrule
    \end{tabularx}
    \caption{Prompts for generating negative instructions for audio and speech. The placeholder \texttt{\{instruction\}} is replaced with the original audio description or speech sentence, while \texttt{\{k\}} denotes the number of negative variants to generate.}
    \label{tab:negative_instruction_prompts}
\end{table*}

\begin{table*}[t]
    \centering

    \footnotesize
    \setlength{\tabcolsep}{7pt}
    \renewcommand{\arraystretch}{1.2}

    \begin{tabularx}{0.95\textwidth}{
        >{\raggedright\arraybackslash\bfseries}p{0.18\textwidth}
        >{\raggedright\arraybackslash}X
    }
        \toprule
        Category & \textbf{Prompt} \\
        \midrule

        QA - \par Audio OVL &
        Listen to the audio. Is it a real and high-quality audio recording? Answer with only one word: yes or no. \\
        \midrule

        QA - \par Speech OVL &
        Listen to the audio. Is it a real and high-quality human voice recording? Answer with only one word: yes or no. \\
        \midrule

        QA - \par Audio REL &
        Does this audio contain the sound events described by the text: \texttt{\{text\}}? Please answer yes or no. \\
        \midrule

        G-Eval - \par Audio OVL &
        Rate how much the audio sounds like a real and high-quality audio recording. Use an integer scale from 1 to 5, where 1 means very poor quality and 5 means excellent quality. Output only one integer: 1, 2, 3, 4, or 5. Do not provide any explanation or additional text. \\
        \midrule

        G-Eval - \par Speech OVL &
        Rate how much the audio sounds like a real and high-quality human voice recording. Use an integer scale from 1 to 5, where 1 means very poor quality and 5 means excellent quality. Output only one integer: 1, 2, 3, 4, or 5. Do not provide any explanation or additional text. \\
        \midrule

        G-Eval - \par Audio REL &
        Rate how closely the audio matches the sound events described by the following text. Use an integer scale from 1 to 5, where 1 means no match and 5 means a perfect match. Output only one integer: 1, 2, 3, 4, or 5. Do not provide any explanation or additional text.

        \par\smallskip
        Text: \texttt{\{instruction\}} \\
        \midrule

        Prompting - \par Audio OVL &
        Rate how much the audio sounds like a real and high-quality audio recording. Use a continuous scale from 1.00 to 5.00, where 1.00 means very poor quality and 5.00 means excellent quality. Output only one number with exactly two decimal places. Do not provide any explanation or additional text. \\
        \midrule

        Prompting - \par Speech OVL &
        Rate how much the audio sounds like a real and high-quality human voice recording. Use a continuous scale from 1.00 to 5.00, where 1.00 means very poor quality and 5.00 means excellent quality. Output only one number with exactly two decimal places. Do not provide any explanation or additional text. \\
        \midrule

        Prompting - \par Audio REL &
        Rate how closely the audio matches the sound events described by the following text. Use a continuous scale from 1.00 to 5.00, where 1.00 means no match and 5.00 means a perfect match. Output only one number with exactly two decimal places. Do not provide any explanation or additional text.

        \par\smallskip
        Text: \texttt{\{text\}} \\

        \bottomrule
    \end{tabularx}
    \caption{Prompts for audio-text relevance, audio/speech quality assessment. The placeholders \texttt{\{text\}} are replaced with the corresponding audio captions.}
    \label{tab:judge_prompts}
    \vspace{100px}
\end{table*}

Tables~\ref{tab:training_audio_prompts}--Table\ref{tab:judge_prompts} list the prompts used in \ours. 
We use the same prompts for UNISON and Audex-2B. Tables~\ref{tab:training_audio_prompts} and~\ref{tab:training_speech_prompts} show the prompts for proposing audio captions and speech transcripts. 
Table~\ref{tab:dialogue_generation_prompts} shows the two-stage prompts for multi-speaker dialogues. 
Table~\ref{tab:training_data_prompts} shows the prompts for generating QAs.
Table~\ref{tab:negative_instruction_prompts} shows the prompts for generating perturbed instructions.
Table~\ref{tab:judge_prompts} lists the judging prompts.

\section{Full Experiment Results on Judge Benchmark}
\label{app:full_judge_results}

\providecommand{\betterbg}[1]{\cellcolor{green!20}#1}
\begin{table*}[t]
\centering
\begingroup
\scriptsize
\setlength{\tabcolsep}{3.0pt}
\renewcommand{\arraystretch}{1.05}
\begin{adjustbox}{width=\textwidth}
\begin{tabular}{l*{6}{c}@{\hspace{8pt}}*{6}{c}@{\hspace{8pt}}*{3}{c}}
\toprule
\textbf{Method}
& \multicolumn{6}{c}{\textbf{Audio-OVL}}
& \multicolumn{6}{c}{\textbf{Speech-OVL}}
& \multicolumn{3}{c}{} \\
\cmidrule(lr){2-7} \cmidrule(lr){8-13}
& \multicolumn{3}{c}{\textbf{PAM}}
& \multicolumn{3}{c}{\textbf{Clotho}}
& \multicolumn{3}{c}{\textbf{BVCC}}
& \multicolumn{3}{c}{\textbf{QualiSpeech}}
& \multicolumn{3}{c}{\textbf{Overall}} \\
\cmidrule(lr){2-4} \cmidrule(lr){5-7} \cmidrule(lr){8-10} \cmidrule(lr){11-13} \cmidrule(l){14-16}
& PCC$\uparrow$ & SRCC$\uparrow$ & MSE$\downarrow$
& PCC$\uparrow$ & SRCC$\uparrow$ & MSE$\downarrow$
& PCC$\uparrow$ & SRCC$\uparrow$ & MSE$\downarrow$
& PCC$\uparrow$ & SRCC$\uparrow$ & MSE$\downarrow$
& PCC$\uparrow$ & SRCC$\uparrow$ & MSE$\downarrow$ \\
\midrule
\multicolumn{16}{c}{\textbf{UNISON}} \\
\midrule
Prompting (greedy) & 0.140 & 0.107 & 1.525 & 0.422 & 0.408 & 0.898 & 0.172 & 0.143 & 3.118 & 0.217 & 0.258 & 2.416 & 0.238 & 0.229 & 1.989 \\
Prompting (avg@8) & 0.198 & 0.178 & 0.873 & 0.393 & 0.359 & \textbf{0.599} & 0.206 & 0.167 & 2.348 & 0.348 & 0.337 & 1.823 & 0.286 & 0.260 & 1.411 \\
AQAScore & 0.242 & 0.227 & 0.684 & \textbf{0.467} & \textbf{0.495} & 0.741 & 0.292 & 0.314 & 1.229 & 0.480 & 0.503 & 1.073 & 0.370 & 0.385 & 0.932 \\
G-Eval & 0.298 & 0.288 & 0.788 & 0.399 & 0.368 & 0.748 & 0.285 & 0.262 & \textbf{0.948} & 0.400 & 0.383 & 1.124 & 0.346 & 0.325 & 0.902 \\
AQAScore + G-Eval & 0.298 & 0.286 & 0.684 & 0.460 & 0.459 & 0.706 & 0.317 & 0.309 & 1.037 & 0.491 & 0.491 & \textbf{1.029} & 0.392 & 0.386 & 0.864 \\
\addlinespace[2pt]
\rowcolor{gray!10}
\ours & \betterbg{\textbf{0.401}} & \betterbg{\underline{0.388}} & \betterbg{\textbf{0.527}} & \underline{0.461} & \underline{0.465} & \underline{0.610} & \betterbg{\underline{0.363}} & \betterbg{\underline{0.371}} & 1.063 & \betterbg{\textbf{0.509}} & \betterbg{\textbf{0.522}} & 1.086 & \betterbg{\textbf{0.434}} & \betterbg{\underline{0.437}} & \betterbg{\textbf{0.821}} \\
\rowcolor{gray!10}
\quad w/o self-merging & \betterbg{\underline{0.399}} & \betterbg{\textbf{0.395}} & \betterbg{\underline{0.588}} & 0.453 & 0.464 & 0.691 & \betterbg{\textbf{0.379}} & \betterbg{\textbf{0.395}} & \underline{1.021} & \betterbg{\underline{0.500}} & \betterbg{\underline{0.510}} & \underline{1.068} & \betterbg{\underline{0.433}} & \betterbg{\textbf{0.441}} & \betterbg{\underline{0.842}} \\
\midrule
\multicolumn{16}{c}{\textbf{Audex-2B}} \\
\midrule
Prompting (greedy) & 0.269 & 0.281 & 1.129 & 0.147 & 0.169 & 0.964 & 0.322 & 0.329 & 1.266 & 0.425 & 0.424 & 1.305 & 0.291 & 0.301 & 1.166 \\
Prompting (avg@8) & 0.245 & 0.244 & \underline{1.029} & 0.167 & 0.193 & \textbf{0.816} & 0.389 & 0.433 & 1.019 & 0.472 & 0.493 & 1.138 & 0.318 & 0.341 & \underline{1.001} \\
AQAScore & 0.372 & \textbf{0.400} & 1.873 & 0.178 & 0.187 & 2.791 & 0.317 & 0.407 & 2.810 & 0.372 & \underline{0.510} & 2.304 & 0.310 & 0.376 & 2.444 \\
G-Eval & 0.231 & 0.259 & 1.105 & 0.021 & 0.045 & 0.972 & 0.560 & 0.563 & \textbf{0.607} & 0.462 & 0.461 & 1.327 & 0.319 & 0.332 & 1.003 \\
AQAScore + G-Eval & \underline{0.373} & \underline{0.396} & 1.059 & 0.153 & 0.173 & 1.183 & 0.453 & 0.519 & 1.190 & 0.444 & 0.485 & 1.248 & 0.356 & 0.393 & 1.170 \\
\addlinespace[2pt]
\rowcolor{gray!10}
\ours & 0.370 & 0.379 & \betterbg{\textbf{0.996}} & \betterbg{\underline{0.179}} & \betterbg{\underline{0.197}} & \underline{0.843} & \betterbg{\underline{0.609}} & \betterbg{\textbf{0.625}} & 0.964 & \betterbg{\underline{0.491}} & 0.507 & \betterbg{\underline{1.040}} & \betterbg{\underline{0.412}} & \betterbg{\underline{0.427}} & \betterbg{\textbf{0.961}} \\
\rowcolor{gray!10}
\quad w/o self-merging & \betterbg{\textbf{0.375}} & 0.382 & 1.163 & \betterbg{\textbf{0.180}} & \betterbg{\textbf{0.198}} & 0.944 & \betterbg{\textbf{0.624}} & \betterbg{\underline{0.622}} & \underline{0.878} & \betterbg{\textbf{0.507}} & \betterbg{\textbf{0.513}} & \betterbg{\textbf{1.022}} & \betterbg{\textbf{0.421}} & \betterbg{\textbf{0.429}} & 1.002 \\
\bottomrule
\end{tabular}
\end{adjustbox}
\endgroup
\caption{Performance on audio / speech quality assessment benchmarks.
Formatting follows Table~\ref{tab:understanding}.}
\label{tab:judge-ovl}
\end{table*}

\providecommand{\betterbg}[1]{\cellcolor{green!20}#1}
\begin{table*}[t]
\centering
\begingroup
\scriptsize
\setlength{\tabcolsep}{3.0pt}
\renewcommand{\arraystretch}{1.05}
\begin{adjustbox}{width=\textwidth}
\begin{tabular}{l*{12}{c}@{\hspace{8pt}}*{3}{c}}
\toprule
\textbf{Method}
& \multicolumn{3}{c}{\textbf{PAM}}
& \multicolumn{3}{c}{\textbf{Clotho}}
& \multicolumn{3}{c}{\textbf{XACLE}}
& \multicolumn{3}{c}{\textbf{RELATE}}
& \multicolumn{3}{c}{\textbf{Overall}} \\
\cmidrule(lr){2-4} \cmidrule(lr){5-7} \cmidrule(lr){8-10} \cmidrule(lr){11-13} \cmidrule(l){14-16}
& PCC$\uparrow$ & SRCC$\uparrow$ & MSE$\downarrow$
& PCC$\uparrow$ & SRCC$\uparrow$ & MSE$\downarrow$
& PCC$\uparrow$ & SRCC$\uparrow$ & MSE$\downarrow$
& PCC$\uparrow$ & SRCC$\uparrow$ & MSE$\downarrow$
& PCC$\uparrow$ & SRCC$\uparrow$ & MSE$\downarrow$ \\
\midrule
\multicolumn{16}{c}{\textbf{UNISON}} \\
\midrule
Prompting (greedy) & 0.373 & 0.307 & 1.426 & 0.445 & 0.409 & 1.044 & 0.419 & 0.338 & 0.877 & 0.353 & 0.288 & 1.397 & 0.398 & 0.336 & 1.186 \\
Prompting (avg@8) & 0.466 & 0.414 & 0.994 & 0.498 & 0.476 & 0.823 & 0.464 & 0.409 & 0.705 & 0.394 & 0.338 & 1.162 & 0.455 & 0.409 & 0.921 \\
AQAScore & 0.569 & \textbf{0.603} & 1.195 & 0.526 & 0.554 & 1.008 & 0.556 & \textbf{0.551} & 0.818 & 0.492 & \textbf{0.534} & 1.398 & 0.536 & \textbf{0.560} & 1.105 \\
G-Eval & 0.494 & 0.452 & 0.637 & 0.530 & 0.500 & 0.595 & 0.508 & 0.461 & 0.636 & 0.413 & 0.369 & 1.019 & 0.486 & 0.446 & 0.722 \\
AQAScore + G-Eval & 0.569 & 0.560 & 0.768 & 0.546 & 0.547 & 0.683 & 0.568 & 0.533 & 0.574 & 0.496 & 0.470 & 1.014 & 0.545 & 0.528 & 0.760 \\
\addlinespace[2pt]
\rowcolor{gray!10}
\ours & \betterbg{\textbf{0.601}} & \underline{0.582} & \betterbg{\textbf{0.580}} & \betterbg{\underline{0.591}} & \betterbg{\underline{0.583}} & \betterbg{\textbf{0.541}} & \betterbg{\textbf{0.591}} & \underline{0.550} & \betterbg{\underline{0.537}} & \betterbg{\textbf{0.515}} & 0.484 & \betterbg{\textbf{0.923}} & \betterbg{\textbf{0.575}} & \underline{0.550} & \betterbg{\textbf{0.645}} \\
\rowcolor{gray!10}
\quad w/o self-merging & \betterbg{\underline{0.590}} & 0.568 & \betterbg{\underline{0.607}} & \betterbg{\textbf{0.595}} & \betterbg{\textbf{0.594}} & \betterbg{\underline{0.545}} & \betterbg{\textbf{0.591}} & 0.545 & \betterbg{\textbf{0.533}} & \betterbg{\underline{0.511}} & \underline{0.486} & \betterbg{\underline{0.927}} & \betterbg{\underline{0.572}} & 0.548 & \betterbg{\underline{0.653}} \\
\midrule
\multicolumn{16}{c}{\textbf{Audex-2B}} \\
\midrule
Prompting (greedy) & 0.252 & 0.229 & 2.808 & 0.315 & 0.338 & 1.798 & 0.230 & 0.211 & 1.658 & 0.198 & 0.178 & 2.411 & 0.249 & 0.239 & 2.169 \\
Prompting (avg@8) & 0.317 & 0.296 & 2.561 & 0.394 & 0.449 & \underline{1.499} & 0.277 & 0.268 & 1.481 & 0.232 & 0.208 & 2.204 & 0.305 & 0.305 & 1.936 \\
AQAScore & 0.312 & \textbf{0.460} & 2.861 & 0.309 & 0.414 & 2.102 & 0.296 & 0.368 & 1.655 & 0.237 & 0.298 & 2.450 & 0.289 & 0.385 & 2.267 \\
G-Eval & 0.355 & 0.388 & \underline{2.285} & 0.434 & 0.458 & \textbf{1.427} & 0.271 & 0.313 & 1.427 & 0.268 & 0.276 & 2.137 & 0.332 & 0.359 & \underline{1.819} \\
AQAScore + G-Eval & 0.389 & \underline{0.442} & 2.512 & 0.410 & 0.468 & 1.690 & 0.335 & 0.367 & 1.479 & 0.295 & 0.321 & 2.246 & 0.357 & 0.400 & 1.982 \\
\addlinespace[2pt]
\rowcolor{gray!10}
\ours & \betterbg{\textbf{0.428}} & 0.415 & \betterbg{\textbf{2.225}} & \betterbg{\textbf{0.471}} & \betterbg{\textbf{0.502}} & 1.687 & \betterbg{\underline{0.407}} & \betterbg{\underline{0.405}} & \betterbg{\textbf{1.268}} & \betterbg{\underline{0.359}} & \betterbg{\underline{0.345}} & \betterbg{\textbf{1.955}} & \betterbg{\textbf{0.416}} & \betterbg{\textbf{0.417}} & \betterbg{\textbf{1.784}} \\
\rowcolor{gray!10}
\quad w/o self-merging & \betterbg{\underline{0.393}} & 0.379 & 2.349 & \betterbg{\underline{0.440}} & \betterbg{\underline{0.473}} & 1.816 & \betterbg{\textbf{0.418}} & \betterbg{\textbf{0.409}} & \betterbg{\underline{1.294}} & \betterbg{\textbf{0.369}} & \betterbg{\textbf{0.359}} & \betterbg{\underline{2.021}} & \betterbg{\underline{0.405}} & \betterbg{\underline{0.405}} & 1.870 \\
\bottomrule
\end{tabular}
\end{adjustbox}
\endgroup
\caption{Performance on audio-text relevance judging benchmarks.
Formatting follows Table~\ref{tab:understanding}.}
\label{tab:judge-rel}
\end{table*}

Tables~\ref{tab:judge-ovl} and~\ref{tab:judge-rel} report the judging results on each benchmark, which are averaged to obtain Table~\ref{tab:judge-overall}.

\end{document}